\documentclass[epj,twocolumn]{webofc}

\usepackage[varg]{txfonts}
\usepackage{physmacros}

\woctitle{LHCP 2013}

\begin{document}

\title{Recent QCD Results From ATLAS}

\author{Christopher Meyer\inst{1}\fnsep\thanks{\email{chris.meyer@cern.ch}} on behalf of the ATLAS Collaboration}

\institute{The Enrico Fermi Institute, The University of Chicago}

\abstract{A survey of recent QCD results using the ATLAS detector at the LHC is presented.}

\maketitle


\section{Introduction}
\label{sec:intro}
The precision measurement of basic quantum chromodynamic (QCD) observables provides information on various aspects of the Standard Model.
Measurements of high-\pt (hard) QCD processes involving jets and photons can be used to constrain the gluon portion of the protons parton distribution functions (PDFs) at high-momentum fraction.
Jet physics also provides a check of the strong coupling constant, \alphas.
The underlying event arising from multiple-parton interactions, beam-beam remnants, and initial/final state radiation provides an irreducible background to all measurements.
As such, a good description by Monte Carlo (MC) simulation is essential for making precision measurements, and searches for physics beyond the standard model.
A measurement of the effective area parameter for double-parton scattering has also been performed, which is an important background in certain searches.
This proceeding to the LHCP 2013 conference provides a brief summary of recent results on QCD using the ATLAS \cite{Aad:2008zzm} detector at the LHC.

\section{Hard QCD}
\label{sec:hardqcd}

\subsection{Jet Physics}
\label{subsec:jet}
\begin{figure}
\centering
\includegraphics[width=7cm,clip]{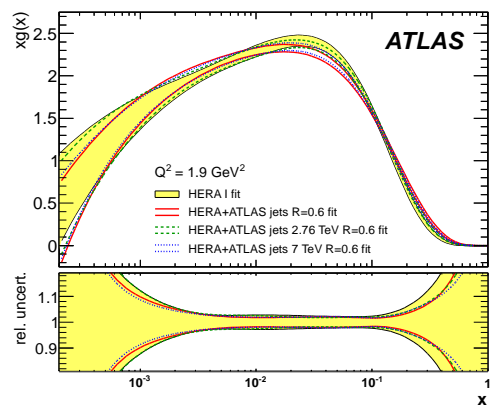}
\caption{
The gluon portion resulting from various PDF fits, including different combinations of HERA-I and ATLAS $2.76\TeV$ and $7\TeV$ data \cite{Aad:2013lpa}.
}
\label{fig:jet_pdffit}
\end{figure}

Inclusive jet cross sections have been measured at $\sqrt{s} = 2.76\TeV$ for anti-$k_t$ jets with $|y| < 4.4$ and \pt up to $300\GeV$ \cite{Aad:2013lpa}.
Because the pileup conditions at $\sqrt{s} = 2.76\TeV$ are similar to those of the 2010 run at $\sqrt{s} = 7\TeV$, the same jet energy calibration is used.
This provides a detailed understanding of the correlations of the jet energy calibration uncertainty between the two measurements.
The double-differential cross section has been measured as a function of both \pt and \y, so that the experimental uncertainties are much reduced when the ratio of $2.76\TeV$ is taken with $7\TeV$.
The measurement is also performed in bins of $\xt = 2\pt/\sqrt{s}$ and \y, where the theoretical uncertainties largely cancel between the two centre-of-mass energies.
A PDF fit exploiting the measurements at both $2.76\TeV$ and $7\TeV$ is performed, providing a strong constraint on the gluon PDF at high-momentum fraction (see figure \ref{fig:jet_pdffit}).

\begin{figure}
\centering
\includegraphics[width=7cm,clip]{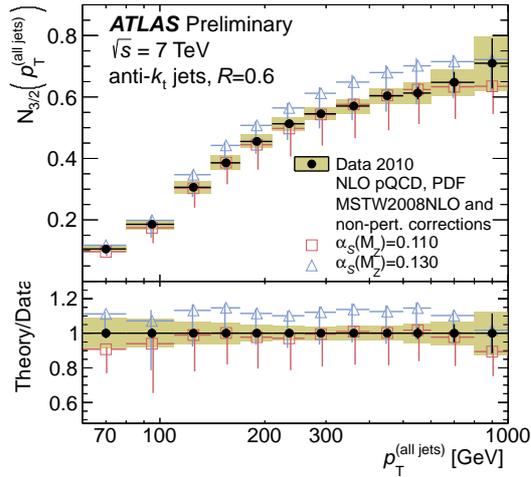}
\caption{
The inclusive jet cross section taken as a ratio for events with $\ge 3$ jets to events with $\ge 2$ jets \cite{ATLAS-CONF-2013-041}.
Theory predictions using \nlojet interfaced with the MSTW 2008 PDF set and including non-perturbative corrections are shown for two separate values of \alphas.
}
\label{fig:multijet_ratio}
\end{figure}

Multijet production \cite{ATLAS-CONF-2013-041} provides a direct probe to the dependence of the theory prediction on higher order terms.
Two observables are defined:
First, the cross section of events with $\ge 3$ jets divided by the cross section of events with $\ge 2$ jets, both as a function of highest jet-\pt.
Second, the ratio of $\ge 3$-jet to $\ge 2$-jet samples of the inclusive jet cross section as a function of jet \pt.
A ratio of the two observables is taken (see figure \ref{fig:multijet_ratio}) to reduce the uncertainty on the jet energy calibration, the dominant source of error.
Because the first definition is proportional to the probability that a two-jet event radiates a third jet (thus is proportional to \alphas), and is less sensitive to the choice of renormalisation/factorisation scale, it is used in the fit for \alphas.
The best fit for \alphas is determined using \nlojet predictions interfaced with the MSTW 2008 PDF set, for a scan of \alphas values.
A best fit value of $\alphas(M_Z) = 0.111 \pm 0.006\mathrm{(exp.)} ^{+0.016}_{-0.003)}\mathrm{(theory)}$ is found, showing good agreement with the global average.
The fit value for \alphas using different \pt bins is evolved to the average \pt value for each bin (up to $800\GeV$) using the two-loop approximation of the Renormalization Group Equation, where agreement within the experimental uncertainties is seen when compared to the world average.

\subsection{Photon Production}
\label{subsec:photon}
The inclusive photon \cite{ATLAS-CONF-2013-022} and diphoton \cite{ATLAS-CONF-2013-023} cross sections measure prompt photon production with minimal surrounding activity.
The cross sections include direct photons (those produced by the hard collision) as well as fragmentation photons (resulting from the fragmentation of a high-\pt parton).
In general, photons in an acceptance $|\eta^{\gamma}| < 1.37$ and $1.52 \le |\eta^{\gamma}| < 2.37$ are used to avoid uninstrumented portions of the electromagnetic calorimeter.
The inclusive photon cross sections as a function of $E_\mathrm{T}$ is well described within uncertainties by next-to-leading order (NLO) theory predictions made by {\sc Jetphox} (which includes both direct and fragmentation contributions).
A slight deficit in the theory prediction is observed for low-$E_\mathrm{T}$, while the data is overestimated by the theory prediction at high-$E_T$.

\begin{figure}
\centering
\includegraphics[width=7cm,clip]{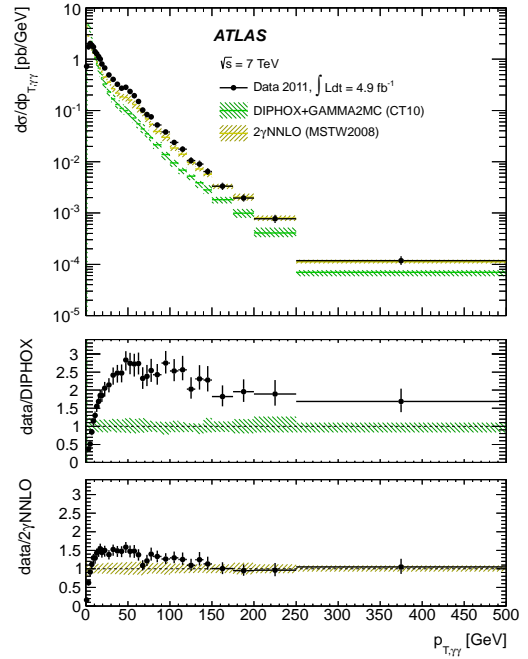}
\caption{
The diphoton cross section as a function of the transverse momentum of the diphoton system \cite{ATLAS-CONF-2013-023}.
The black points are data, the green bands are the DIPHOX prediction, and the yellow band is the 2$\gamma$NNLO prediction.
}
\label{fig:diphoton_e}
\end{figure}

The \pt of two-photon systems is compared to theory predictions by DIPHOX and 2$\gamma$NNLO.
DIPHOX includes both direct and fragmentation components at NLO, as well as the NNLO diagram for $gg \to \gamma \gamma$.
2$\gamma$NNLO includes the full NNLO prediction of the direct photon contribution, however neglects the fragmentation component.
The NNLO prediction best describes data, except at low \pt where the fragmentation contribution is large (see figure \ref{fig:diphoton_e}).

\begin{figure}
\centering
\includegraphics[width=7cm,clip]{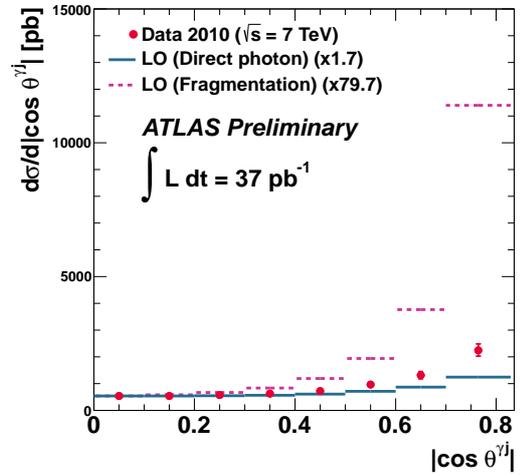}
\caption{
The cross section for photon production in association with a jet \cite{Aad:2012tba}.
The solid pink circles are data, the lines represent the leading order prediction for the direct photon contribution (blue) and the fragmentation contribution (pink).
}
\label{fig:photonjet_costheta}
\end{figure}

Measuring photon production in association with a jet \cite{Aad:2012tba} provides an interesting probe of $|\cos \theta^{\gamma j}|$, which is sensitive to the spin of the exchange particle.
Good agreement is observed compared with the predictions of {\sc Jetphox}, using multiple PDF sets.
The angular distribution also serves as a discriminating variable between photons produced directly and by fragmentation, as seen in figure \ref{fig:photonjet_costheta}.

\section{Underlying Event}
\label{sec:ue}

\subsection{Event Shape}
\label{subsec:ue}
\begin{figure}
\centering
\includegraphics[width=7cm,clip]{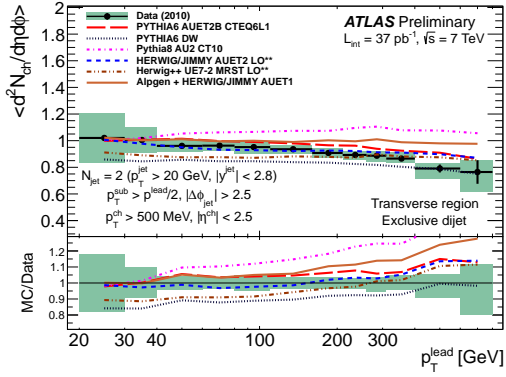}
\caption{
The number of charged particles per unit area ($\eta/\phi$) in the transverse region, for events where only two jets are present \cite{ATLAS-CONF-2012-164}.
Data (solid black points) are compared with leading order MC predictions.
}
\label{fig:ue_nch}
\end{figure}

A measurement of the underlying event has been performed in ATLAS which focuses on inclusive jet and dijet events \cite{ATLAS-CONF-2012-164}, considering jets of $\pt > 20\GeV$ and $|y|<2.8$.
Distributions of charged particle multiplicity, charged and inclusive $\sum\pt$ densities, and mean charged-particle \pt are studied in the ``transverse region,'' defined as the region $\pi/3 \le |\Delta \phi| < 2\pi/3$ from the highest-\pt jet in the event.
Activity in the transverse region is increased due to NLO emission, such that the difference in activity between two transverse regions is also an interesting observable.
Good agreement is observed when restricting the comparison of data and leading order MC simulation to events with exactly two jets, as expected in a region of phase space with little emission.
In general the MC simulation shows decent agreement across a variety of variables, with \herwig performing slightly better describing the properties of charged particles in underlying event (see figure \ref{fig:ue_nch}).

\subsection{Double-parton Scattering}
\label{subsec:mpi}
\begin{figure}
\centering
\includegraphics[width=7cm,clip]{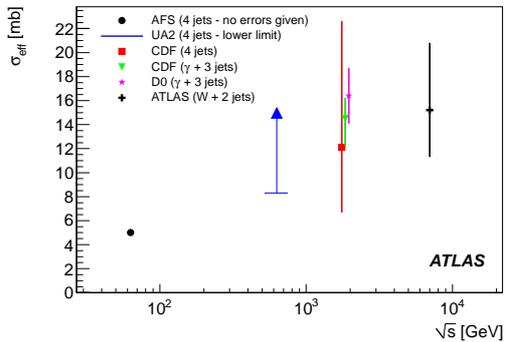}
\caption{
The effective cross section for double parton scattering in ATLAS, compared with previous results \cite{Aad:2013bjm}.
}
\label{fig:mpi_sigma}
\end{figure}

At higher $\sqrt{s}$ the low momentum-fraction region where PDFs are large is probed, so that multiple-parton contributions can become non-negligible.
This gives rise to an important background for many single parton scattering measurements.
The ATLAS analysis of double-parton scattering \cite{Aad:2013bjm} employs a template fit to determine the fraction of events where a $W$ is produced in association with exactly two jets arising from double-parton interactions.
Jets with $\pt > 20\GeV$ and $|y|<2.8$ are considered for this measurement.
The double-parton production fraction is used to derive the effective area parameter ($\sigma_{\mathrm{eff}}$) for hard double-parton scattering.
As shown in figure \ref{fig:mpi_sigma} the result of $\sigma_{\mathrm{eff}} = 15 \pm 3 (\mathrm{stat.}) ^{+5}_{-3} (\mathrm{sys.})$ mb is consistent with those measured by previous experiments.

\bibliography{LHCP2013-Meyer-QCD}

\end{document}